%% file: main.tex
\documentclass{svproc}
\usepackage{url}
\usepackage[super]{nth}
\newcommand{\rom}[1]{\uppercase\expandafter{\romannumeral #1\relax}}
\usepackage{todonotes}

\usepackage{todonotes}

\usepackage[ruled,vlined]{algorithm2e}
\usepackage{amssymb}
\usepackage{subcaption}
\usepackage{bm}
\usepackage{multirow}
\usepackage{comment}

\begin{document}
\mainmatter              
\title{Detecting Anomalies in Software Execution Logs with Siamese Network}
\titlerunning{Detecting Anomalies in Software Execution Logs with Siamese Network}  
%
\author{Shayan Hashemi\inst{1} \and Mika Mäntylä\inst{1}}
\authorrunning{Shayan Hashemi et al.} 
%
%
\institute{M3S Research Unit, ITEE, University of Oulu, Finland\\
\email{shayan.hashemi@oulu.fi},
\email{mika.mantyla@oulu.fi}
}

\maketitle              

\begin{abstract}
  \input{texts/abstract}
\end{abstract}

\section{Introduction}
    \input{texts/intro}
\section{Background and Related Works}

    \subsection{General Log Analyzer Architecture}
        \input{texts/background}
        
    \subsection{Related Works}
        \input{texts/related_works}
        
    \subsection{Datasets}
        \input{texts/datasets}

\section{Proposed Method}
    \input{texts/proposed_method}

\section{Data and Performance measures}
    \input{texts/experiments}

\section{Conclusion and Future Work}
    \input{texts/conclusion}

\section{Acknowledgment}
    \input{texts/acknowledgements}
    
\bibliographystyle{ieeetr}
\bibliography{refs}

\end{document}

%% file: texts/abstract.tex
Logs are semi-structured text files that represent software's execution paths and states during its run-time. Therefore, detecting anomalies in software logs reflect anomalies in the software's execution path or state. So, it has become a notable concern in software engineering. We use LSTM like many prior works, and on top of LSTM, we propose a novel anomaly detection approach based on the Siamese network. This paper also provides an authentic validation of the approach on the Hadoop Distributed File System (HDFS) log dataset. To the best of our knowledge, the proposed approach outperforms other methods on the same dataset at the $F_1$ score of $0.996$, resulting in a new state-of-the-art performance on the dataset.
Along with the primary method, we introduce a novel training pair generation algorithm that reduces generated training pairs by the factor of $3000$ while maintaining the $F_1$ score, merely a modest decay from 0.996 to 0.995. Additionally, we propose a hybrid model by combining the Siamese network with a traditional feedforward neural network to make end-to-end training possible, reducing engineering effort in setting up a deep-learning-based log anomaly detector.
Furthermore, we examine our method's robustness to log evolutions by evaluating the model on synthetically evolved log sequences; we got the $F_1$ score of 0.95 at the noise ratio of $20\%$. Finally, we dive deep into some of the side benefits of the Siamese network. Accordingly, we introduce a method of monitoring the evolutions of logs without label requirements at run-time. Additionally, we present a visualization technique that facilitates human administrations of log anomaly detection.

%% file: texts/intro.tex

Log files are an unstructured text-based history of events that shed light on the software state during its execution. Each line of log files indicates a different event and may hold different types of information such as log-type, timestamp, process ID, thread ID, and log message. Analyzing log events allows developers to extract helpful information from the software state during the run-time. One of the log analysis applications is anomaly detection. Log anomaly detection may assist developers in software testing, debugging, or run-time monitoring.

Recently, deep learning has become the most predominant method in almost every machine learning problem. Furthermore, deep neural networks have been utilized to improve software testing, debugging, and stability. Going more in-depth, DNNs are used in applications such as software defect prediction \cite{dnndefect}, performance analysis \cite{dnnperf}, or reopened bugs accuracy prediction \cite{dnnbugpred}. Moreover, log anomaly detection is no exception, and DNNs have been widely utilized in this research area alongside other Machine Learning (ML) approaches.

There are two different approaches among the deep methods in log anomaly detection \cite{anodetsurvey}. The first one is a binary classification task. It takes a sequence as input and outputs a binary value indicating if the sequence is an anomaly. The latter approach is sequence modeling, which trains only on the non-anomaly data, learns to model the system's normal behavior resulting in predictions of low probabilities for anomaly behavior.

 As non-anomaly data volume is significantly higher than anomaly data, sequence modeling is more common in log anomaly detection. However, training solely on non-anomaly data may result in models being unaware of anomaly events, making the approach unreliable in anomaly situation. Furthermore, since logs evolve due to software updates, models trained with non-anomaly data have limited capabilities to detect anomaly situations in evolved non-anomaly situations. 
 
On the other hand, binary classification solves the previously mentioned problem by training the model on both anomaly and non-anomaly data. However, it comes with its own challenges; one of them is training on an unbalanced dataset. The obstacle comes into place when the proportion of anomaly to non-anomaly data is too small. More specifically, datasets contain dramatically more anomaly samples in comparison to non-anomaly ones.

Nonetheless, many solutions have been introduced to surmount the unbalanced data obstacle. Oversampling and undersampling are two straightforward approaches that strive to equalize the number of samples in two classes. Another way of dealing with unbalanced datasets is weighted training. It manipulates the cost function so that both classes' influences on the model's parameters are equal. However, setting training weights and oversampling may result in overfitting, while undersampling ignores a colossal proportion of negative samples during the training process. A more steady solution may be synthetic data generation. Furthermore, it eliminates the disadvantages of oversampling yet results in equilibrium. However, it requires innovative methods to generate legitimate and reliable samples. This paper proposes a new approach based on the Siamese network \cite{siamese} to handle the unbalanced data in log anomaly detection.

The primary purpose of the Siamese network is similarity learning, and it is vastly used in one-shot learning such as face verification \cite{siameseface1,siameseface2}, signature verification \cite{siamesesig1,siamesesig2}, and visual object tracking \cite{siameseobjtrack1,siameseobjtrack2,siameseobjtrack3}. Furthermore, it has also been proposed in the context of anomaly detection in video games \cite{wilkins2020anomaly}. The proposed Siamese-network-based model takes advantage of both non-anomaly and anomaly data while not demanding balanced training data.

More in-depth, we attempt to learn an embedding function for log sequences that maps sequences of the same class (non-anomaly or anomaly) adjacent to each other while maximizing the distance between opposing classes' sequences. We also propose a sampling technique inspired by negative sampling \cite{negsampling} to generate pairs for the Siamese network's training process. The proposed algorithm significantly reduces the training costs of the Siamese network.

Furthermore, we evaluate the proposed method through various experiments. Accordingly, we examine the impact of different pair generation algorithms on the Siamese network, try different classifiers on top of the embedding neural network, and compare the best performer to state-of-the-art methods. Moreover, we evaluate our model's robustness on evolved log sequences and propose a method to monitor log evolutions at production time. Besides, we reveal a solution to visualize the embedded sequences to make human administration of log sequences possible. Finally, we construct a hybrid model by imposing the Siamese network on a feedforward neural network, investigating the Siamese network's positive impact.

The remainder of the paper is organized as follow:
Section \rom{2} is dedicated to explaining required knowledge, reviewing famous previous works, and discussing datasets. The Siamese network, the methodology, and pair generation algorithms are explained in Section \rom{3}. The preprocessing, dataset, and evaluation metric are discussed in Section \rom{4}.  Section \rom{5} comprises the reports of various experiments investigating the proposed method on deeper levels, while additional practical advantages are mentioned in Section \rom{6}. Finally, conclusion and future work proposals are offered in Section \rom{7}. 

%% file: texts/background.tex
Log anomaly detection consists of multiple components, which are visualized in Figure \ref{figure_system_architecture}. The figure illuminates four components that each Log anomaly detector holds: preprocessor, log parser, log vectorizer, and classifier.
\input{figures/system_architecture}

The first component is the preprocessor. As its name implies, its mission is to prepare log events for subsequent components. The preparations may include eliminating unnecessary information (such as IP addresses or invalid characters), extracting features from timestamps and log levels, and clustering logs based on their threads or process IDs. The preprocessor unit's output is passed to the next component, the Log parser \cite{loghubtools}. The log parser identifies the log message parameters and extracts templates. Log message event types could be inferred by matching a log message with identified templates. Depending on the vectorizer's capabilities, which is the next component, event parameters might be carried along with the event type. The log vectorizer produces vectors from event types and parameters (if any). The vectors may take the form of one-hot encoded, semantic, or template IDs depending on the classifier's architecture. Then, vectors are given to the classifier, which is the last component. The classifier's goal is to distinguish anomalous vectors. Machine learning algorithms are quite prevalent for this component, as they have shown promising results in sequence modeling and classification.

%% file: figures/system_architecture.tex
\begin{figure}
  \includegraphics[width=\textwidth]{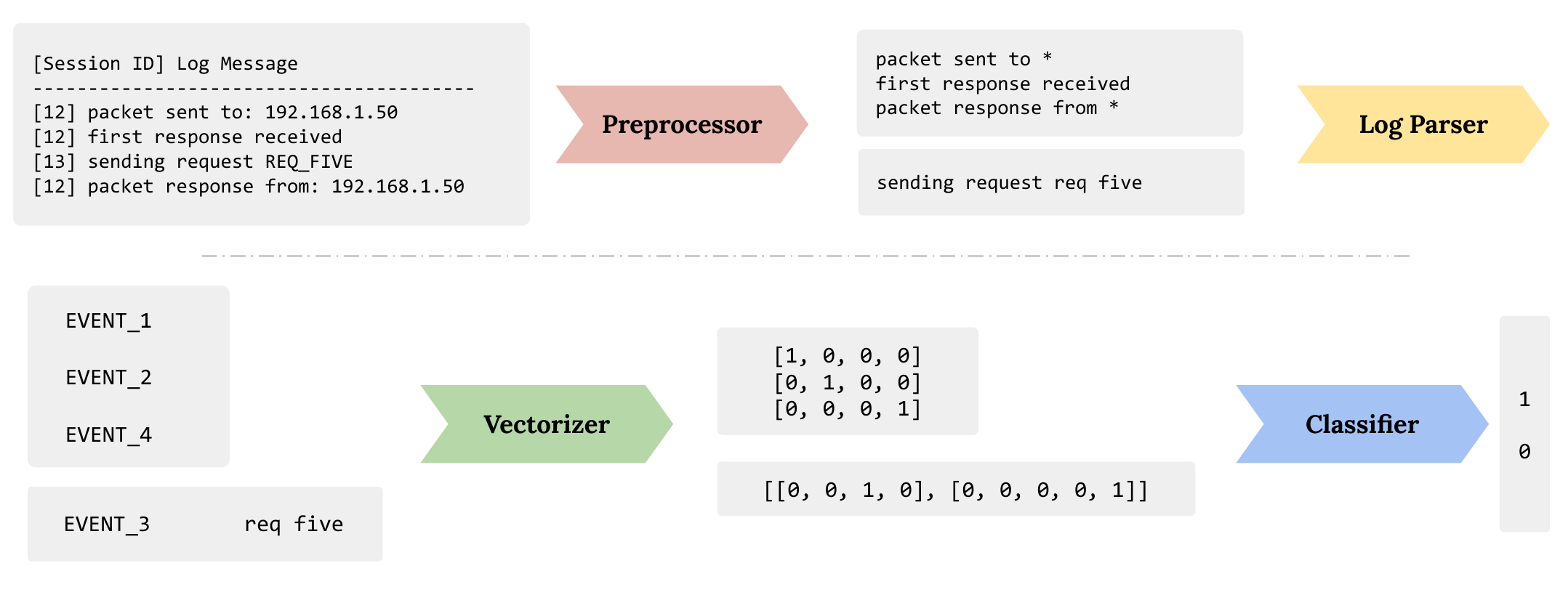}
  \caption{System log anomaly detector's architecture as in \cite{logrobust}.}
  \label{figure_system_architecture}
\end{figure}

%% file: texts/related_works.tex
One of the most well-known and effective log anomaly detection methods is PCA based mentioned in \cite{pca}. The method first forms a session-event matrix, similar to the document-term matrix in Natural Language Processing (NLP), where each cell indicates the number of occurrences of a particular event that occurred in an individual session. Next, the matrix is passed to an analysis of principal components. Then the anomalies are detected by distinguishing the session vector's projection length in the residual space. 

In another approach, \cite{invarmining} uses the session-event matrix and mine invariants that satisfy the majority of the sessions. Thus, anomalies occur in sessions that lack the satisfaction of the mined invariants. While all mentioned works focus on designing general-purpose algorithms, \cite{cloudseer} presents a method that compares the log messages to a set of automata to calculate the workflow divergence and is labeled as an anomaly as a result. However, it focuses on the log anomaly detection in OpenStack's logs specifically.

As the Deep Neural Networks have grown more mature in recent years, they have gained popularity among log anomaly detection research. Many approaches are leveraging different types of Recurrent Neural Networks (RNNs) such as Long Short-Term Memory (LSTM) \cite{lstm} or Gated Recurrent Unit (GRU) \cite{gru}, while others are detecting anomalies by making use of Convolutional Neural Networks (CNNs) \cite{cnn}.

DeepLog, as the most well-known log anomaly detection method, proposed in \cite{deeplog}, uses DNN in the classifier component. After parsing log events, DeepLog encodes the event types and parameters into vectors. Next, the model, which is based on LSTM, trains on data from non-anomaly execution only to predict the next log event given previous events. After the training, the model predicts a low probability for some events in anomaly sequences as it has trained on non-anomaly data only. 

Although the methods mentioned before accurately detect log anomalies, \cite{logrobust} suggests that advances made by previous works are based on a close(d)-world assumption where logs are static, while, in real-world applications, logs are continuously evolving. Log evolutions are considered undoubtedly important these days, as many companies are continuously delivering software updates to their customers \cite{leppanen2015highways}. Thus, the authors of \cite{logrobust} suggest LogRobust, a novel method for log anomaly detection. LogRobust proposes a new vectorization technique called ``semantic vectorization'' to approximately compensates for the evolution of log messages. It also suggests utilizing the attention-based Bidirectional Long Short-Term Memory (Bi-LSTM) to encounter the execution path evolutions. Furthermore, the authors present a technique to emulate log evolutions by applying noise.

LogAnomaly, explained in \cite{loganomaly}, presents another novel yet practical approach for vectorization called ``template2vec'' that takes synonyms and antonyms into account, making the vectorization process more reliable. Furthermore, LogAnomaly claims that it can detect sequential anomalies as well as quantitive ones. While every previously mentioned deep method applies LSTM to model log sequences (predict the next log event), LogAnomaly uses an LSTM on Term Frequency-Inverse Document Frequency (TF-IDF) vectors to construct a binary classifier. On the other end of the spectrum, \cite{cnnlog} applies CNN instead of LSTM to form a binary classifier. The research also introduces an effective embedding method to transform one-hot encoded log events to vectors called ``log-key2vec''. This method results in efficient dimension reduction of one-hot encoded vectors.

All previous deep-learning-based methods, regardless of their core components, obeyed one of the two previously mentioned approaches. They either applied binary classification or modeled the sequence. However, this paper presents a third option that utilizes the Siamese network to circumvent the previously mentioned challenges in a different matter. Harnessing the Siamese network's power, our method proposes a new approach to embed the log sequences into vectors, so that embedded sequence vectors of different classes are readily separable and classifiable in the new space.

%% file: texts/datasets.tex
In the area of log anomaly detection, many datasets exist. The LogHub data collection contains currently 16 different software log datasets \cite{loghub}. The LogHub collection offers log datasets from various software types such as distributed file systems, operating systems, and web-based services. Additionally, six of them are labeled for the task of anomaly detection.

The labeled logs may be divided into two different categories based on their labeling approach. The first is the one-to-one labeled datasets, consisting of log sequences with a unique label for every log element. In the second category, named n-to-one, on the other hand, there is one label for each sequence of elements. It could also be interpreted as if all elements of an individual log sequence possess the same label (anomaly or non-anomaly).

Among the LogPai's labeled datasets, to the best of our knowledge, the Hadoop Distributed File System (HDFS) and Openstack log datasets for anomaly detection, mentioned in \cite{hdfs,deeplog}, are the only n-to-one labeled datasets suitable for our research. Moreover, since almost all previously mentioned works utilize the labeled datasets on sequence level to detect software log anomalies regardless of their approaches (binary classification or sequence modeling), these two datasets are prevalent among the related works.
As this paper demands one-to-n labeled datasets, we strived to utilize the same datasets (HDFS and Openstack) from the previous works. However, during our tests, after parsing the Openstack using the Drain log parsing algorithm, mentioned in \cite{drain}, we noticed that it lacks a sufficient amount of unique sequences. More in-depth, we found only 11 unique sequences in the Openstack dataset while it was 18,383 in the HDFS dataset. Hence, experiments are limited to the HDFS dataset only, which is also the case in some of the prior works such as \cite{cnnlog}.

%% file: texts/proposed_method.tex
As earlier mentioned, previous deep methods either train on non-anomaly events only or apply binary classification to detect anomalies. However, both of those approaches are prone to deficiencies.

In the first (non-anomaly events only) approach, the model training would not encounter log events that only occur in an anomaly situation. For instance, In a distributed data storage solution software, a hard drive failure event is not a regular event by any means. For example, in the HDFS dataset, from the twenty-nine unique events, only twenty-two of them occurred in non-anomaly situations. Needless to say that not training on a proportion of the input space may result in unexpected model behavior. In the latter (binary classification) approach, the model's training suffers from the unbalanced dataset. Although some solutions have been discussed for the unbalanced data problem, all of them are accompanied by their limitations.

Throughout the rest of this section, we propose a novel approach based on the Siamese networks due to their excellent performance in one-shot learning problems \cite{siameseface1,siameseface2,siameseobjtrack1,siameseobjtrack2,siameseobjtrack3} and their stability on unbalanced data \cite{siameseunbalanced}. Our proposed method takes advantage of both data classes without any sampling tricks or weighted training.

\subsection{The Siamese network}
\input{figures/siamese_architecture}
The Siamese network, illuminated in Figure \ref{figure_siamese_architecture}, was initially invented to resolve the one-shot learning problem \cite{siamese} by forming a similarity-based embedding function. It packs two neural networks with shared weights (they are indeed the same neural networks and may be considered one; however, discriminating them makes the Siamese network's architecture more interpretable) and a similarity metric. During the training, at first, pairs of samples are passed to the neural networks. Next, the neural network embeds them into vectors. Then, the similarities between the vectors are measured. Lastly, the optimization process updates the weights of the neural networks with respect to the fact that similar pairs (same class) should hold high similarity values for their output vectors, while it is the contrary for dissimilar pairs (pair from different classes). At the end of the training process, the model embeds the same class samples close to each other while different class samples are embedded away from each other. In this paper, we use the Siamese network to train a deep embedding neural network that transforms log sequences into vectors so that embedded vectors of sequences of the same class are close to each other while being apart from the other class.

After the Siamese network converges, we extract the embedding neural network and embed all training sequences into vectors. As the embedded vectors of different classes are well separated, they are excellent training data for an arbitrary classifier. So, we train a classifier to work on top of the embedding neural network to form an anomaly detection method. During the test time, the embedding neural network transforms the input sequences into vectors and passes them to the classifier to be classified as non-anomaly or anomaly sequences.

Since the invention of the Siamese network, different loss function has emerged for it. One of them is the contrastive loss function, mentioned in \cite{contrastiveloss}. It operates utilizing the Euclidean distance, confirming enough space between embedded vectors of different classes while keeping vectors from the same class close to each other. However, during our experiments, we inquired about another loss function based on the sigmoid of inner product and cross-entropy loss function \cite{deeplearning}, which performed better than the contrastive loss. Going more in-depth, we use the sigmoid function on embedded sequences' inner product to construct a similarity measure. This measure may be formulated as:
$$
sim(x_1, x_2) = \sigma(x_1\boldsymbol{\cdot} x_2).
$$
On top of the similarity measure, we use the cross-entropy loss function. So, the final loss function may be formulated as:
$$
J(x_1, x_2, y) = -(y.\log{(sim(x_1,x_2))} + (1-y).\log{(1-sim(x_1,x_2))}).
$$

\subsection{Pair generation}
As the Siamese network requires its training input to be in pairs, a proper pair generation method is required. Generated training pairs must include two types of pairs in order to train the Siamese network. The first type is similar pairs in which the entities are from the same class, with the training target being one. The second type is dissimilar pairs in which the entities are from different classes, with the training target set to zero. To shed more light, assume that $A$ is an anomaly sequence, while $N$ is a non-anomaly sequence. From four possible pair permutations, $(A, A)$ and $(N, N)$ are considered as similar pairs, while $(A, N)$ and $(N, A)$ are dissimilar ones. The following paragraphs contain two pair generation algorithms for training the Siamese network.

\input{algorithms/all}
The first approach, which is quite straightforward, generates every possible pair. Going more in-depth, every sequence in the dataset pairs with all other sequences except for itself. The pseudo-code could be seen in Algorithm \ref{algorithm_all}. Although this method is sensible and easy to implement, it is impractical for massive datasets. Alongside the exponential growth of pairs quantity, this approach generates dramatically more similar pairs than dissimilar ones. We call this approach the ``All'' pair generation algorithm.

\input{algorithms/kpone}
The second approach focuses on training efficiency. In this approach, for each sequence within the dataset, we sample one sequence from the same class and $K$ sequences from the different class, generating $K+1$ pairs for each sequence. In other words, this approach samples a subset of all pairs instead of generating them all. The pseudo-code is observable in Algorithm \ref{algorithm_kpone}. This method reduces training time and power consumption, making it feasible for training the Siamese network. We name this approach the "K Plus One (KPOne)" pair generation algorithm. As the $K$ value increases, so does the computational effort. We noticed improvements in our experiments while increasing $K$ until $K=3$.

The number of samples generated in each epoch, and the computational cost accordingly, may vary significantly based on the choice of the pair generation algorithm. Assuming that $n_n$ and $n_a$ are subsequently the number of non-anomaly and anomaly samples within a dataset. The number of pairs generated by the All algorithm is
$$N_{All}=n_a^2 + n_n^2 + 2 n_a n_n - n_a - n_n$$
, while the number of generated pairs for the KPOne algorithm is
$$N_{KPOne}=K n_a + K n_n + n_a + n_n$$
when $K$ is the dissimilar samples count. It is blindingly obvious that for large numbers of $n_a$ and $n_n$, the value of $N_{sample}$ is dramatically smaller than $N_{all}$. So, the computational cost of the All pair generation algorithm is larger than the KPOne.

%% file: figures/siamese_architecture.tex
\begin{figure}
    \centering
    \resizebox{\columnwidth}{!}{
    \includegraphics[width=\textwidth]{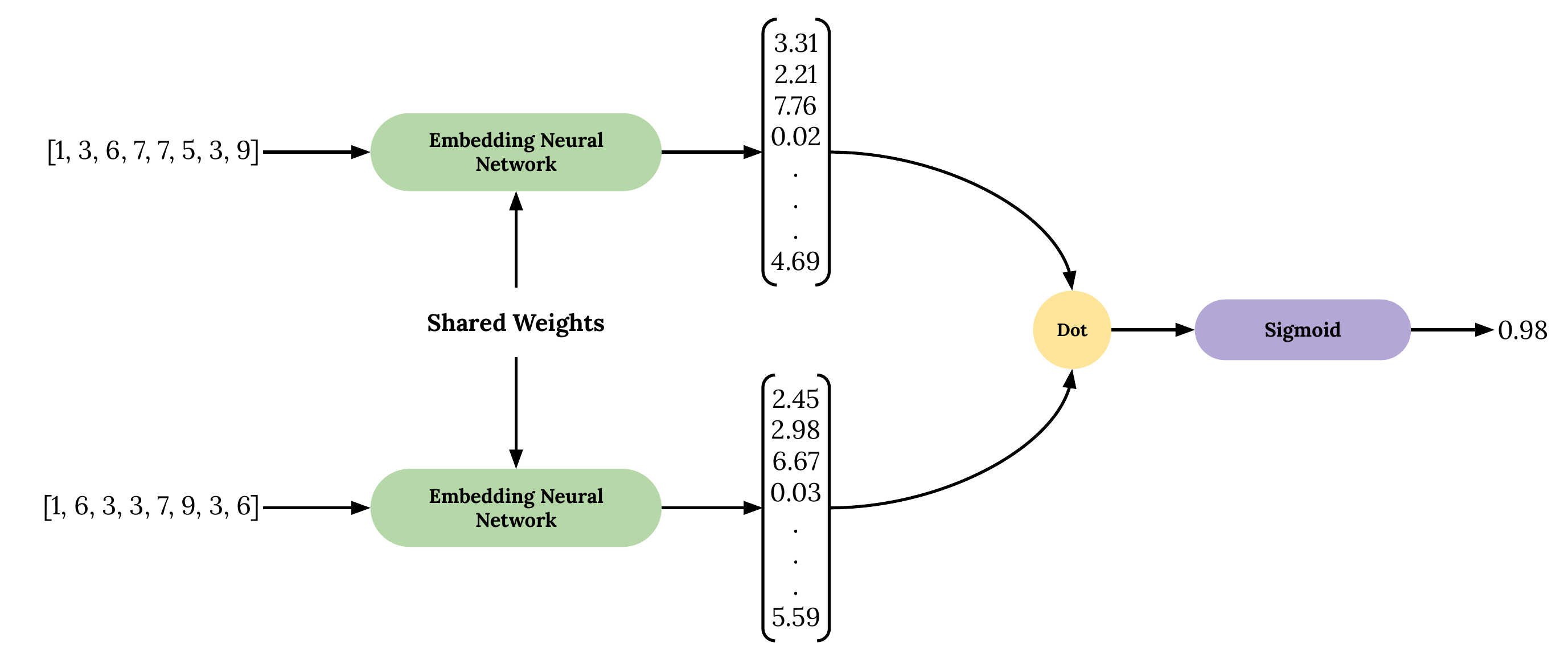}
    }
    \caption{The Siamese network's architecture.}
    \label{figure_siamese_architecture}
\end{figure}

%% file: algorithms/all.tex
\begin{algorithm}[t]
    \caption{Generating pairs using the All algorithm}
    \label{algorithm_all}
    \SetKwInOut{Input}{inputs}
    \SetKwInOut{Output}{output}
    \SetKwProg{GenerateAllPairs}{GenerateAllPairs}{}{}
    \SetKwFunction{AddPair}{addPair}

    \GenerateAllPairs{($D$)}{%
        
        \Input{The dataset $D$, which contains sequences denoted by $s$ and targets denoted by $t$}
        
        \Output{Pairs generated using the All algorithm}
        
        \ForEach{$(s_1, t_1) \in D$}{%
            \ForEach{$(s_2, t_2) \in D$}{%
                \If{$t_1 == t_2$}{%
                    \AddPair{$s_1, s_2, 1$}
                }
                \If{$t_1 != t_2$}{%
                    \AddPair{$s_1, s_2, 0$}
                }

            }
        }
    }
\end{algorithm}

%% file: algorithms/kpone.tex
\begin{algorithm}[t]
    \caption{Generating pairs using the KPOne algorithm}
    \label{algorithm_kpone}
    \SetKwInOut{Input}{inputs}
    \SetKwInOut{Output}{output}
    \SetKwProg{GenerateKPOnePairs}{GenerateKPOnePairs}{}{}
    \SetKwFunction{AddPair}{addPair}
    \SetKwFunction{SampleSet}{sampleSet}

    \GenerateKPOnePairs{($N, P, K$)}{%
        
        \Input{The data subsets $N$ and $P$, which subsequently contain negative (non-anomaly) and positive (anomaly) sequences. The constant $K$ where $K \in \mathbb{N}$ and is the proportion of dissimilar to similar pairs.
        }
        
        \Output{Pairs generated using the KPOne algorithm}
        
        \ForEach{$n \in N$}{%
            sn = \SampleSet{$N$}\;
            \AddPair{$n, sn, 1$}\;
            \For{$0$ \KwTo $K$}{
                sp = \SampleSet{$N$}\;
                \AddPair{$n, sp, 0$}\;
            }
        }
        
        \ForEach{$p \in P$}{%
            sp = \SampleSet{$P$}\;
            \AddPair{$p, sp, 1$}\;
            \For{$0$ \KwTo $K$}{
                sn = \SampleSet{$N$}\;
                \AddPair{$p, sn, 0$}\;
            }
        }
        

    }
\end{algorithm}

%% file: texts/experiments.tex
In this section, we explain the dataset and the preprocessing steps and evaluate our architecture. 

\subsection{Dataset and preprocessing}
We use the HDFS log dataset for anomaly detection as it has become a benchmark in the log anomaly detection task. It is widely used in previous works, making a fair and comprehensive comparison between our and the existing state-of-the-art methods possible. As our research focuses on classification, we use the dataset's vectorized variant, provided by \cite{loghub}, since the input data is cleaned, processed, transformed into sequences, and prepared for classification. The dataset contains 575,061 sequences, 558,223 of them being non-anomaly sequences, while 16,838 are anomaly sequences.

Although the dataset is supposed to be ready for classification, we discovered many redundant sequences. Redundancy not only raises the required processing power for training but also compromises the authenticity of the evaluation as some test samples may appear in the training set. So, our first and only step of preprocessing is to remove redundant sequences. After removing the redundant sequences, the dataset contains 4,124 unique anomaly and 14,259 unique non-anomaly sequences.

\subsection{Data splitting}

We split the data into the train and the test sets (90\% for training and 10\% for testing). The train set is used for training the Siamese network and classifiers, while the test set is utilized for evaluating the system. However, before we start to generate pairs using the desired pair generation algorithm and train the Siamese network, we take a small proportion of training data (equal to 3\% of all data), generate all available pairs from it, and use it as the validation set. Then we start the training using pairs generated with the selected algorithm. The validation set's purpose is to find the most suitable neural network architecture and hyper-parameters and control overfitting. After founding proper architecture and hyper-parameters, the validation set serves no purpose. Thus, it is merged into the training set for retraining the neural network. Figure \ref{figure_overall} illuminates the data splitting and experiments processes, presenting an overall view of the whole process.
\input{figures/overall}


\subsection{Performance measure}
The nature of the anomaly detection task is unbalanced, meaning that there are significantly more negative samples in comparison to positive ones. In such circumstances, the binary classification accuracy is not a valid metric for measuring performance. So, we use another metric called ``$F_1$ score'' to measure and compare performance. Suppose $TP$, $TN$, $FP$, and $FN$ are respectively true positives, true negatives, false positives, and false negatives. The ``precision'' metric  formulated as
$$precision= \frac{TP}{TP+TN}$$
 shows the accuracy of the model's positive prediction. On the other hand, the ``recall'' metric demonstrates the model's reliability in predicting all positive samples and formulates as
$$recall= \frac{TP}{TP+FN}$$
Finally, the $F_1$ score is the harmonic mean of precision and recall simplified to
$$F_1=2\cdot\frac{precision\cdot recall}{precision+recall}$$
However, we multiply $F_1$ scores by one hundred to expose more details in the results.


\section{Experiments, Results, and Comparisons}
This section focuses on spotting a proper architecture for embedding neural network, validating different pair generation algorithms, comparing different classifiers and other state-of-the-art methods, and introducing low-cost and hybrid models.

\subsection{Embedding neural network's architecture}
\textbf{Motivation:}
As our method's heart is the embedding neural network trained inside the Siamese network, we want the embedding neural network to perform at its best. Spotting an optimal architecture and hyper-parameters is a challenging step in deep learning projects. So, we need an algorithm to find suitable architecture and hyper-parameters.

\textbf{Method:}
Multiple algorithms, such as Grid Search, Random Search, Bayesian Optimization, and Evolutionary Optimization, have been proposed for tuning neural network architecture and hyper-parameters. However, we choose the Hyperband algorithm \cite{hyperband} for its performance and computational efficiency to attain a solid architecture and hyper-parameters. The Hyperband algorithm was executed three times (to avoid local optima) with default parameters on all available pairs in training set to minimize the Siamese loss on the validation set. In other words, Hyperband used the training set to train multiple different architectures and the validation set to compare the architectures to find the best performance.

\textbf{Findings:}
Table \ref{table_embedding_architecture} contains the details of the embedding neural network's architecture and hyper-parameters found by the Hyperband algorithm. It shows that multiple layers of LSTMs are required to achieve decent results as sequences in the HDFS log for anomaly detection dataset are quite complicated.
\input{tables/embedding_neural_network_architecture}

\subsection{Pair generation algorithms comparison}
\textbf{Motivation:}
As discussed before, generating pairs using the All pair generation algorithm is computationally expensive. Therefore, we proposed an algorithm for generating pairs to reduce the computational cost. In this experiment, we aim to compare two different algorithms for pair generation.

\textbf{Method:}
We trained two models with the same architecture found in the previous experiment. One trains on pairs generated using the All pair generation algorithm, while the other one's training pairs are generated using the KPOne algorithm. We tried different values for $K$ in the KPOne pair generation algorithm and found out that $k=3$ works the best in our use case. After the training, we compare the Siamese network's loss value and the classifiers' accuracy across the two models. It must be stated that the test loss value is calculated after the hyper-parameter optimization process in the previous section. In fact, the Hyperband algorithm neither trained on nor targeted any pairs containing any sequence from the test set.

\textbf{Findings:}
The results, available in Table \ref{table_siamse_loss}, show that the All pair generation algorithm results in less error in the Siamese network.  However, Table \ref{table_classifiers_results} (in the next subsection) demonstrates that the classification result differences are negligible. All in all, considering the computational cost (more than 3,000 times generated pairs), the All algorithm might not be a fitting choice for many cases.
\input{tables/siamese_network_error}


\subsection{Classifiers comparison}
\textbf{Motivation:}
After training the embedding neural network inside the Siamese network, a classifier is needed to classify the embedded sequences. In this experiment, we aim to evaluate different classifiers for this purpose.

\textbf{Method:}
We pick Logistic Regression (LR), Support Vector Machine (SVM), K Nearest Neighbours (KNN), and multi-layer neural network as the candidate classifiers. The neural network classifier consists of two layers. The first one is activated using the Rectifier Linear Unit (ReLU), while the second layer leverages the sigmoid activation function for binary classification. We embed all train sequences into vectors and train the classifiers on them. During the test time, each sequence is embedded using the embedding neural network and passed to the classifier for prediction.

\textbf{Findings:}
As Table \ref{table_classifiers_results} exposes the results, all classifiers achieve outstanding results, the $F_1$ score of 99.39 or better. Achieving accurate and consistent results with different classifiers explains that the embedding neural network works precisely and as expected. Since the multi-layer neural network performs better than the other classifiers, we choose it as our Best performer for upcoming experiments.
\input{tables/classifiers_results}

\subsection{Comparison to state-of-the-art methods}
\label{subsection_accuracy_test}
\textbf{Motivation:}
This section evaluates the Best performer from the previous subsection against state-of-the-art deep log anomaly detection approaches.

\textbf{Method:}
We bring the results of the best performers from the previous experiments and select DeepLog \cite{deeplog}, LogRobust \cite{logrobust}, LogAnomaly \cite{loganomaly}, and CNNLog \cite{cnnlog} as the competitors. We also train a neural network with the architecture equal to combining the embedding and classifier neural networks as a single unit. This neural network, called the Feedforward model, allows us to investigate if utilizing the Siamese network yields any benefit.

\textbf{Findings}:
Table \ref{table_other_methods} shows that our Best performer outperforms all previous works and its Feedforward rival. We see that our Best performer has the $F_1$ score of 99.62 followed by LogRobust with the $F_1$ score of 99\footnote{The source gives no decimals given so the actual $F_1$ score value could be anything between 98.50 and 99.49)}, CNNLog ($F_1$ score 98.5), the Feedforward model ($F_1$ score 97.28), DeepLog ($F_1$ score 96), and LogAnomaly ($F_1$ score 95). Nevertheless, to the best of our knowledge, our Best performer achieves the best results ever on the HDFS log for anomaly detection data set yet. Moreover, the Siamese network outperforming its Feedforward rival shows that applying the Siamese network results in an increase in the $F_1$ score.
\input{tables/other_methods_comparison}

\subsection{Low-cost model}
\textbf{Motivation:}
In previous experiments, we found an architecture offering the state of the art performance for anomaly detection in the HDFS dataset.  
However, training a model with that architecture is expensive and was done in HPC-environment. In this experiment, we endeavor to handcraft a new architecture that is less taxing to train. After all, the software industry might not have the possibility or time to train models in HPC-environment. Furthermore, experiments, development, and utilization are cheaper and faster for the low-cost model. Finally, as the low-cost model is capable of computationally more efficient inferences, it demands less computational power, making it economical, fast, and scalable at the production time. However, despite all benefits, the low-cost model sacrifices accuracy to achieve them.

\textbf{Method:}
With the goal to find a suitable architecture, we first handcraft different architectures that are significantly less expensive to train than the architecture found by the Hyperband. Later, we train all models using the KPOne pair generation algorithm with $k=3$. In the end, we choose the best architecture according to the $F_1$ score. Alongside the $F_1$ score, we record two different metrics for both models. The first metric is the number of floating-point operations (FLOPS) for one forward pass of the neural network. FLOPS is an implicit indication of computational cost during both development and production. Additionally, we calculate the number of parameters for each model. The number of parameters specifies the amount of memory required to store and load the model and explicitly affects the training speed. Finally, we compare training time in a typical deep learning machine's hardware (A 14 cores Intel Xeon CPU with Nvidia Tesla P100 GPU).

\textbf{Findings:}
Table \ref{table_handcrafted_architecture} demonstrates the chosen handcrafted architecture. Table \ref{table_handcrated_results} compares the Best Performer model and low-cost model in computational cost, model size, and accuracy. The comparison sheds light on the fact that despite being computationally more affordable, three times less floating-point operation, 30 times fewer parameters, and reducing the training time by the factor of 13, the low-cost architecture does not considerably compromise the $F_1$ score, from 99.62 to 98.78. For example, the low-cost model could be retrained overnight with typical hardware while it is not possible for the best performer in typical hardware. This would make it suitable for environments where logs would evolve rapidly, but less accuracy is tolerated.   
\input{tables/handcrafted_architecture}
\input{tables/handcrafted_results}

\subsection{Hybrid Model - Combining the Siamese and Feedforward networks}
\textbf{Motivation:}
As previous experiments indicate, the best performer architecture is the classifier neural network on top of the embedding neural network. However, since the classifier and the embedding function are both neural networks, we strive to train them together, making end-to-end training possible. The end-to-end architecture may reduce design and engineering efforts as the classifier and embedding neural networks train simultaneously. 

\textbf{Method:}
Before training, we place the classifier network after the last component of the embedding neural network in the Siamese network. Therefore, the modified Siamese network is going to have two outputs. The first one is the similarity indicator, while the second one is the predicted label for the first entry of the Siamese network. Therefore, the modified Siamese network's loss is the cumulative loss of the Siamese network and cross-entropy classification. Figure \ref{figure_end_to_end_architecture} visualizes the architecture the modified Siamese network. Furthermore, to analyze the impact of the Siamese network on the accuracy, we compare the Hybrid model with another model with the same architecture without the Siamese similarity loss, i.e., the Feedforward model mentioned in \ref{subsection_accuracy_test}.
\input{figures/end_to_end_architecture}

\textbf{Findings:}
Table \ref{table_end_to_end_standalone} confirms that the Hybrid model performs better than the Feedforward model and is almost on a par with separately trained embedding and classifier neural networks (the Best performer), both performing at the $F_1$ score of 0.99.
\input{tables/end_to_end_standalone}


\section{Practical Advantages}
This section notes some practical advantages that become possible with the Siamese network. The first two advantages are related to log evolution, and the last one is related to log visualization. 

\subsection{Robustness}
\textbf{Motivation:}
Software logs are continually evolving due to different execution environments or developers' updates. Moreover, \cite{logrobust} performed an empirical study confirming how software logs evolve. As training deep learning models is dramatically power consuming, it is not feasible to train the model for every minor software updates or changes in execution environments. Accordingly, \cite{logrobust} introduces three methods for emulating log evolution synthetically by adding noise to log sequences. It is not rational to train models on synthetically generated data. However,  synthetically generated data may help in evaluating and analyzing model performance on evolved logs.

\textbf{Method:}
In this experiment, we decided to apply the three methods of adding noise to log sequences, mentioned in \cite{logrobust}, to imitate the evolutions of log sequences. The methods comprise of duplicating, removing, and shuffling one or multiple elements of a sequence. Since generating a noisy dataset is a random process, we ran each test five times and took the results' average. 

\textbf{Findings}:
 Table \ref{table_noisy_test} shows the classifiers' evaluation of synthetically evolved log sequences with different noise ratios. Harnessing the power of the Siamese network, classifiers maintained their accuracy formidably despite the evolutions. The $F_1$ score drops from 0.99 to 0.92 in all classifiers when moving from noise ratio of 0\% to noise ratio of 30\%. In previous works, with the sequence of noise ratio of 20\%, the $F_1$ score of 0.95 was observed. However, direct comparison is not possible since the noise parameters were not fully disclosed, and our setup might differ from their experiments. Nevertheless, showing the $F_1$ score of 0.95 at 20\% noise ratio, in Table \ref{table_noisy_test}, our method's robustness appears to be as strong as previous works.
\input{tables/noisy_sequence_test}

\subsection{Unsupervised log evolution monitoring}
\textbf{Motivation:}
We have confirmed that the proposed model is considerably robust to log sequence evolution in the previous section. However,  if log sequences proceed to evolve, the retraining process is inevitable. Since the retraining process is computationally expensive and time-consuming, we strive to find a solution to minimize the number of retraining times. More in-depth, we seek a numeral value to present the trained model's reliability on evolved sequences. Although the $F_1$ score accuracy is the best measurement for reliability, we do not possess the sequence labels to calculate the $F_1$ score in the production time as the incoming data is completely new. It goes without mentioning that it is possible to label the production data later down the road, yet labels are not available at the exact moment of production.
Hence, we require a new metric that indicates the reliability without any labeling requirement.

\textbf{Method:}
Since the embedding neural network transforms sequences into vectors, we may exploit embedded vectors' distribution to monitor log sequences' evolution. Thus, we introduce the fitness score as the indication of evolutions in log sequences.
To calculate the fitness score, the training sequences are embedded into vectors using the embedding neural network and modeled by fitting into a Gaussian mixture. Accordingly, the fitness score is computed as the average log-likelihood of embedded vectors of evolved sequences. The more the log sequences evolve, the lower the fitness value will be. Possessing such a metric, we may define a threshold and avoid the retraining process for trivial evolutions in production. Moreover, we may retrain the model as soon as the fitness score surpassed the threshold number. Needless to say, the threshold number might vary from task to task or even dataset to dataset.

\textbf{Findings:}
We used the previously mentioned methods to imitate log evolutions and recorded the fitness score as the evolutions increased. Figure \ref{figure_distribution_drift} visualizes the purge in fitness score as the evolutions grow. The purge might be an indication of the fitness score's reliability.
\input{figures/distribution_drift}

\subsection{Sequence visualization}
\textbf{Motivation:}
We have proposed multiple methods of evaluating the authenticity and reliability of the embedding neural network and model in previous experiments. However, human supervision for AI systems can bring brighter insights. One of the best solutions to human supervision is visualization. Furthermore, the visualization (of the embedding neural network's output in our case) gives humans the ability to supervise the embedding neural network's output and allows manual analysis.

\textbf{Method:}
As the trained embedding network allows us to transform log sequences into vectors, we can use dimension reduction algorithms such as T-SNE \cite{tsne}, UMAP \cite{umap}, and PCA \cite{principalca} to reduce the dimensions of the embedded sequences so that they become visualizable and perceptible for humans. Accordingly, we embed all sequences from the train and test sets to vectors, reduce their dimensions, and plot the results on a canvas.

\textbf{Findings:}
Figure \ref{figure_visualization} visualizes the embedded sequences using different dimension reduction methods. The embedded non-anomaly sequences are colored as blue, while the anomaly ones are colored as red. The figure demonstrates that embedded sequences of different classes (non-anomaly/anomaly) are readily separable regardless of the dimension reduction algorithm. This fact might explain the high accuracy among all the classifiers.
\input{figures/vectors_visualization}

%% file: figures/overall.tex
\begin{figure}
  \includegraphics[width=\textwidth]{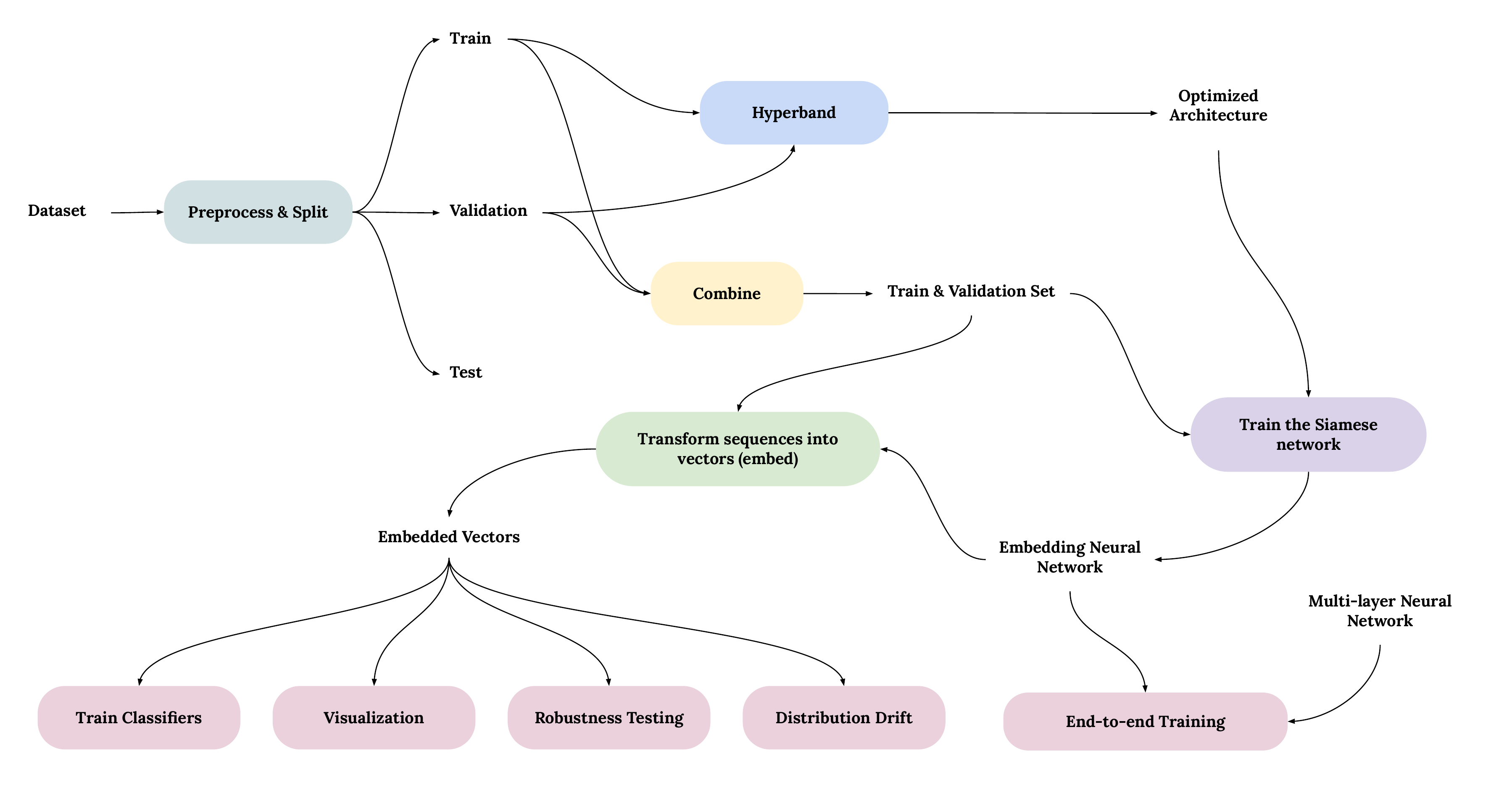}
  \caption{Overall view of data splitting and experiments.}
  \label{figure_overall}
\end{figure}

%% file: tables/embedding_neural_network_architecture.tex
\begin{table}
  \caption{The embedding neural network's architecture found by the Hyperband algorithm, described layer by layer.}
  \label{table_embedding_architecture}
  \begin{center}
  \begin{tabular}{ll@{\hskip 6pt}l@{\hskip 6pt}l@{\hskip 6pt}l@{\hskip 6pt}l@{\hskip 6pt}l@{\hskip 6pt}l}
        \hline%
    \multicolumn{1}{l}{\rule{0pt}{12pt}Property}        & \nth{1}    & \nth{2}  & \nth{3}  & \nth{4}  & \nth{5}  & \nth{6}  & \nth{7} \\ [2pt]
        \hline\rule{0pt}{12pt}%
    Layer Type    & Embedding & LSTM  & LSTM  & LSTM & Dense & Dense & Dense\\
    Output Units  & 11        & 192   & 192   & 64   & 348   & 640   & 64   \\
    Activation    & N.A       & Tanh  & Tanh  & Tanh & ReLU  & ReLU  & Linear \\ [2pt]
        \hline
    \end{tabular}
    \end{center}
\end{table}

%% file: tables/siamese_network_error.tex
\begin{table}
  \caption{The Siamese network's loss using different pair generation algorithms.}
  \label{table_siamse_loss}
    \begin{center}
        \begin{tabular}{lrrr}
            \hline
            \rule{0pt}{12pt}Algorithm & Train Loss & Test Loss & Generated Pairs\\[2pt]
                \hline\rule{0pt}{12pt}%
            All                      & $\sim$0.00 & 0.0025 & 211,231,337  \\
            KPOne                   & 0.01       & 0.03 & 63,968     \\[2pt]
            \hline
        \end{tabular}
    \end{center}
\end{table}

%% file: tables/classifiers_results.tex
\begin{table}
  \caption{The accuracy comparison between different classifiers and embedding neural network trained using different pair generation algorithm. The first model is trained using the All pair generation algorithm while the second one is trained using the KPOne.}
  \label{table_classifiers_results}
  \begin{center}
    \begin{tabular}{lrr}
        \hline
    \multirow{2}{*}{Classifier}          & \multicolumn{2}{c}{\rule{0pt}{12pt} $F_1$ Score} \\[2pt]
               & All Algorithm  & KPOne Algorithm \\[2pt]
        \hline\rule{0pt}{12pt}%
            K Nearest Neighbours   & 99.39                                                         & 99.39                                                         \\
            Support Vector Machine & 99.57                                                         & 99.51                                                         \\
            Neural Network         & 99.62                                                         & 99.51                                                         \\
            Logistic Regression    & 99.39                                                         & 99.39 \\ [2pt]
        \hline
        \end{tabular}
    \end{center}
\end{table}

%% file: tables/other_methods_comparison.tex
\begin{table}
  \caption{The comparison of the Best performer from our approaches and other state-of-the-art deep methods. It should be noted that the numbers in this table are not multiplied by one hundred. Since reported numbers in other works were not as accurate as ours, we could not demonstrate more precise metrics for them.}

  \label{table_other_methods}
  \begin{center}
    \begin{tabular}{lrrr}
        \hline
        \rule{0pt}{12pt}Method              & Precision & Recall & $F_1$ Score \\[2pt]
        \hline\rule{0pt}{12pt}%
        DeepLog \cite{deeplog}            & 0.95     & 0.96  & 0.96      \\
        LogRobust \cite{logrobust}         & 0.98    & 1.00  & 0.99       \\
        LogAnomaly \cite{loganomaly}        & 0.96     & 0.94  & 0.95       \\
        CNNLog \cite{cnnlog}            & 0.977     & 0.993  & 0.985       \\
        Best performer     & 0.9931     & 0.9994  & 0.9962       \\
        Feedforward model & 0.9924     & 0.9539  & 0.9728      \\[2pt]
        \hline
    \end{tabular}
  \end{center}
\end{table}

%% file: tables/handcrafted_architecture.tex
\begin{table}
  \caption{The handcrafted embedding neural network's architecture found by cross-validation between ten different candidate models.}
  \label{table_handcrafted_architecture}
  \begin{center}
  \begin{tabular}{ll@{\hskip 6pt}l@{\hskip 6pt}l@{\hskip 6pt}l@{\hskip 6pt}l}
        \hline%
    \multicolumn{1}{l}{\rule{0pt}{12pt}Property}        & \nth{1}    & \nth{2}  & \nth{3}  & \nth{4}  & \nth{5} \\ [2pt]
        \hline\rule{0pt}{12pt}%
    Layer Type    & Embedding & Bi-LSTM  & Dense  & Dense & Dense\\
    Output Units  & 24        & 64 ($32\times2$)   & 64   & 64   & 64\\
    Activation    & N.A       & Tanh  & Leaky ReLU  & Leaky ReLU & Linear\\ [2pt]
        \hline
    \end{tabular}
    \end{center}
\end{table}

%% file: tables/handcrafted_results.tex
\begin{table}
  \caption{The table is the comparison of low-cost architecture with the architecture found by Hyperband. FLOPS column indicates the amount of floating-point operations required for the embedding neural network to transform a sequence into a vector. Moreover, the Parameters column reveals the number of trainable parameters in each architecture. Furthermore, the required training time for each architecture is mentioned in the Training time column.}

  \label{table_handcrated_results}
  \begin{center}
    \begin{tabular}{lrrrrr}
        \hline
        \rule{0pt}{12pt}Architecture & $F_1$ Score & FLOPS & Parameters & Training time\\[2pt]
        \hline\rule{0pt}{12pt}%
        Best performer            & 99.62     & 222K  & 805K & 150h 42min\\
        Low-cost         & 98.78     & 71K   & 27K & 11h 17min\\[2pt]
        \hline
    \end{tabular}
  \end{center}
\end{table}

%% file: figures/end_to_end_architecture.tex
\begin{figure}
    \centering
    \resizebox{\columnwidth}{!}{
    \includegraphics[]{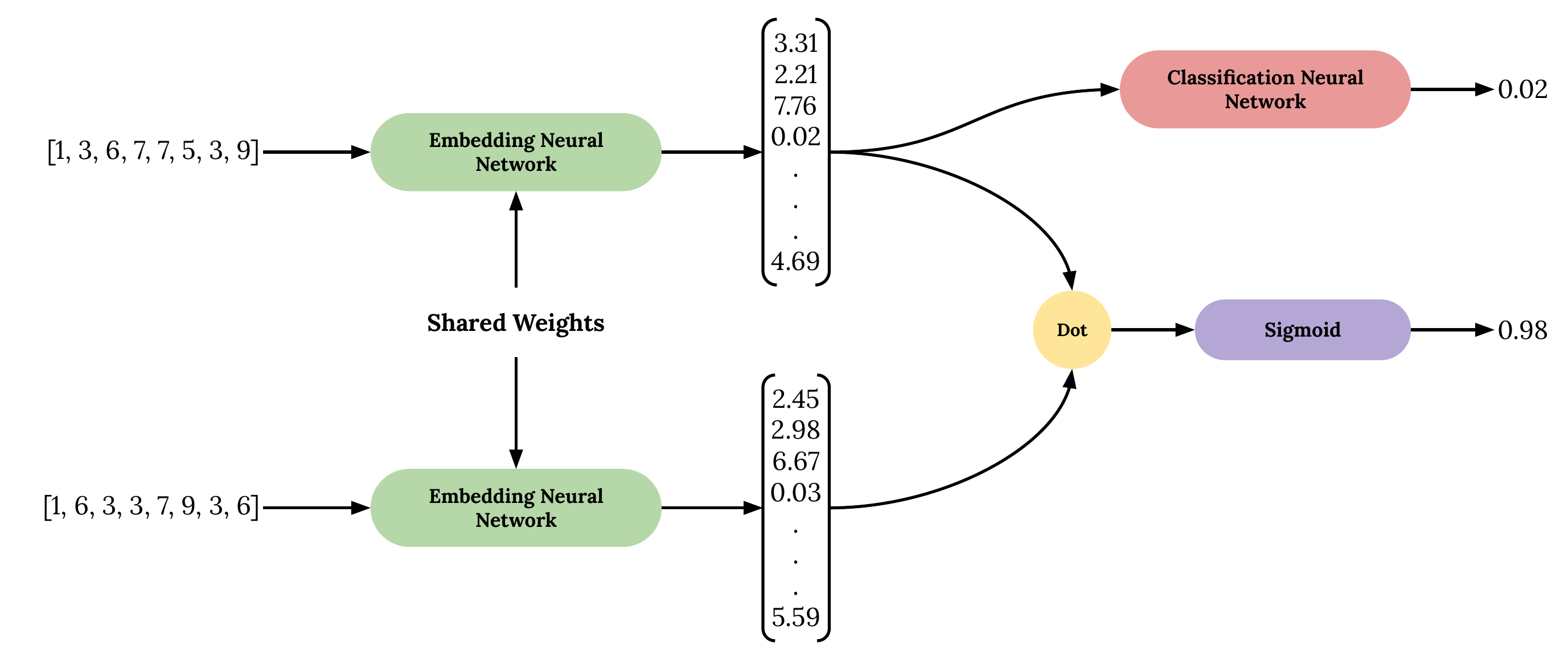}
    }
    \caption{The modified Siamese network's architecture for end-to-end training.}
    \label{figure_end_to_end_architecture}
\end{figure}

%% file: tables/end_to_end_standalone.tex
\begin{table}
    \caption{The comparison of end-to-end training model (training classifier alongside the embedding neural network) with the best performer and feedforward model.}
    \label{table_end_to_end_standalone}
    \begin{center}
        \begin{tabular}{lrrr}
        \hline%
        \rule{0pt}{12pt} Model            & Precision & Recall & $F_1$ Score \\
        \hline\rule{0pt}{12pt}%
        Hybrid Model & 0.9927     & 0.9878  & 0.9902        \\
        Best performer     & 0.9931     & 0.9994  & 0.9962       \\
        Feedforward model & 0.9924     & 0.9539  & 0.9728      \\[2pt]
        \hline
        \end{tabular}
    \end{center}
\end{table}

%% file: tables/noisy_sequence_test.tex
\begin{table}
    \caption{The evaluation results of different classifiers on synthetically evolved datasets. The noise ration indicates the ratio of the test set samples that are affected by synthetic log evolutions.}
    \label{table_noisy_test}
    \begin{center}
        \begin{tabular}{llrrrrr}
            \hline
                \multirow{2}{*}{Classifier} & \multirow{2}{*}{Metric}             & \multicolumn{5}{c}{\rule{0pt}{12pt}Noise Ratio} \\
                                                            & &      0\%      & 5\%    & 10\%   & 20\%   & 30\%  \\[2pt]

            \hline\rule{0pt}{12pt}%
                \multirow{3}{*}{K Nearest Neighbors}   & Precision      & 99.76 & 97.78  & 96.47  & 94.02  & 90.87 \\
                                                        & Recall        & 99.03 & 98.16  & 97.68  & 96.61  & 94.77 \\
                                                        & $F_1$ score   & 99.39 & 97.97  & 97.07  & 95.30  & 92.77 \\
        \rule{0pt}{12pt}%
                \multirow{3}{*}{Support Vector Machine} & Precision         & 99.76 & 97.83  & 96.56  & 93.98  & 90.76 \\
                                                        & Recall            & 99.03 & 98.16  & 97.72  & 96.71  & 94.92 \\
                                                        & $F_1$ score       & 99.39 & 97.99  & 97.14  & 95.32  & 92.78 \\
            \rule{0pt}{12pt}%
                \multirow{3}{*}{Neural Network}         & Precision     & 99.31 & 97.69  & 96.52  & 93.67  & 90.51 \\
                                                        & Recall        & 99.94 & 97.72  & 97.72  & 96.71  & 94.86 \\
                                                        & $F_1$ score   & 99.62 & 97.92  & 97.11  & 95.16  & 92.68 \\
            \rule{0pt}{12pt}%
                \multirow{3}{*}{Logistic Regression}    & Precision     & 99.76 & 97.02  & 96.65  & 94.33  & 91.58 \\
                                                        & Recall        & 99.03 & 98.11  & 97.58  & 96.51  & 94.62 \\
                                                        & $F_1$ score   & 99.39 & 98.02  & 97.11  & 95.86  & 93.07 \\ [2pt]
            \hline
        \end{tabular}
    \end{center}

\end{table}

%% file: figures/distribution_drift.tex
\begin{figure}
    \centering
  \includegraphics[width=\linewidth]{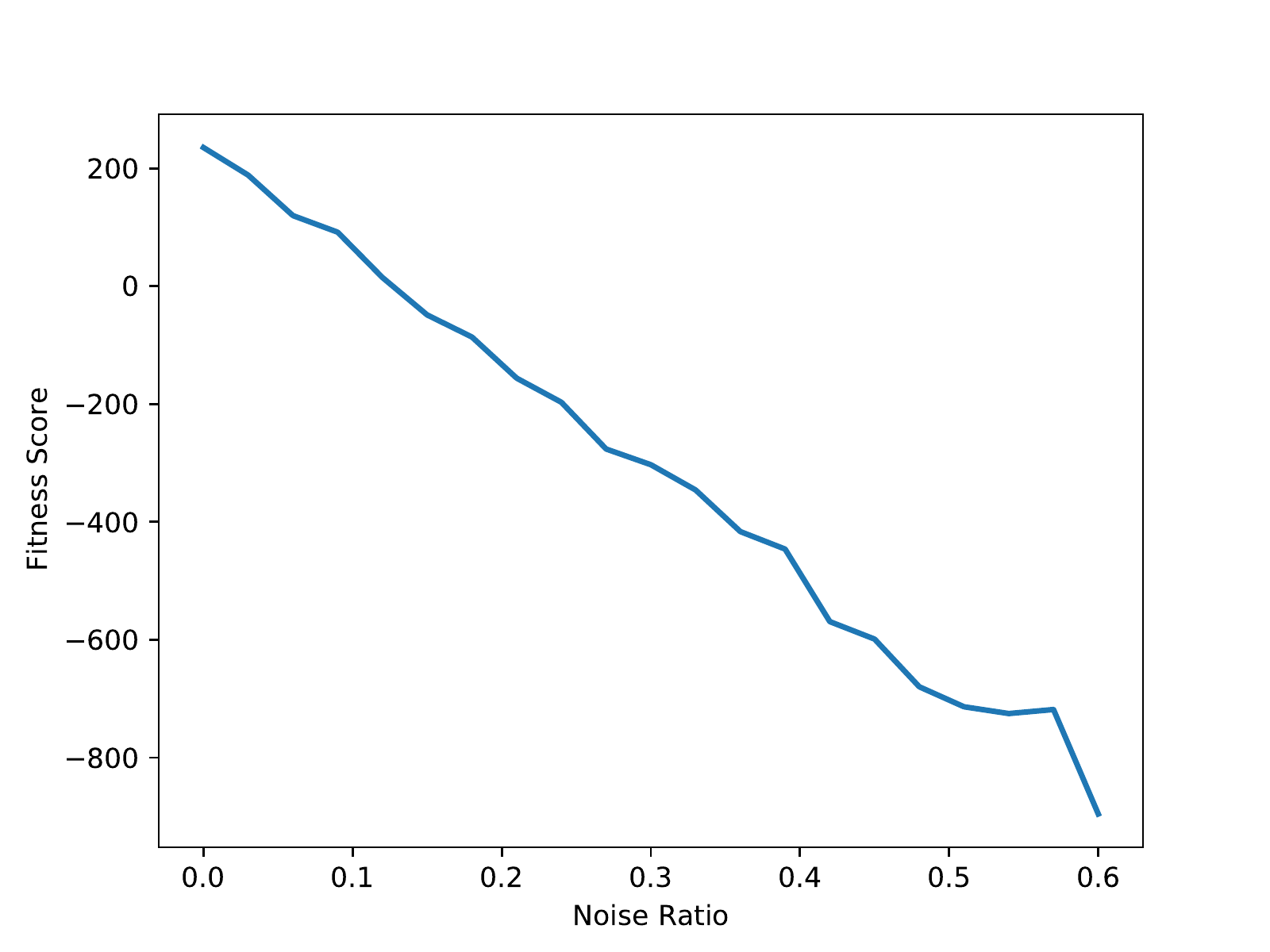}
  \caption{The purge in fitness score as the noise ratio increase. It should be noted that positive scores are due to computing scores using probability density function.}
  \label{figure_distribution_drift}
\end{figure}

%% file: figures/vectors_visualization.tex
\begin{figure}

  \begin{subfigure}[b]{0.32\linewidth}
    \includegraphics[width=\linewidth]{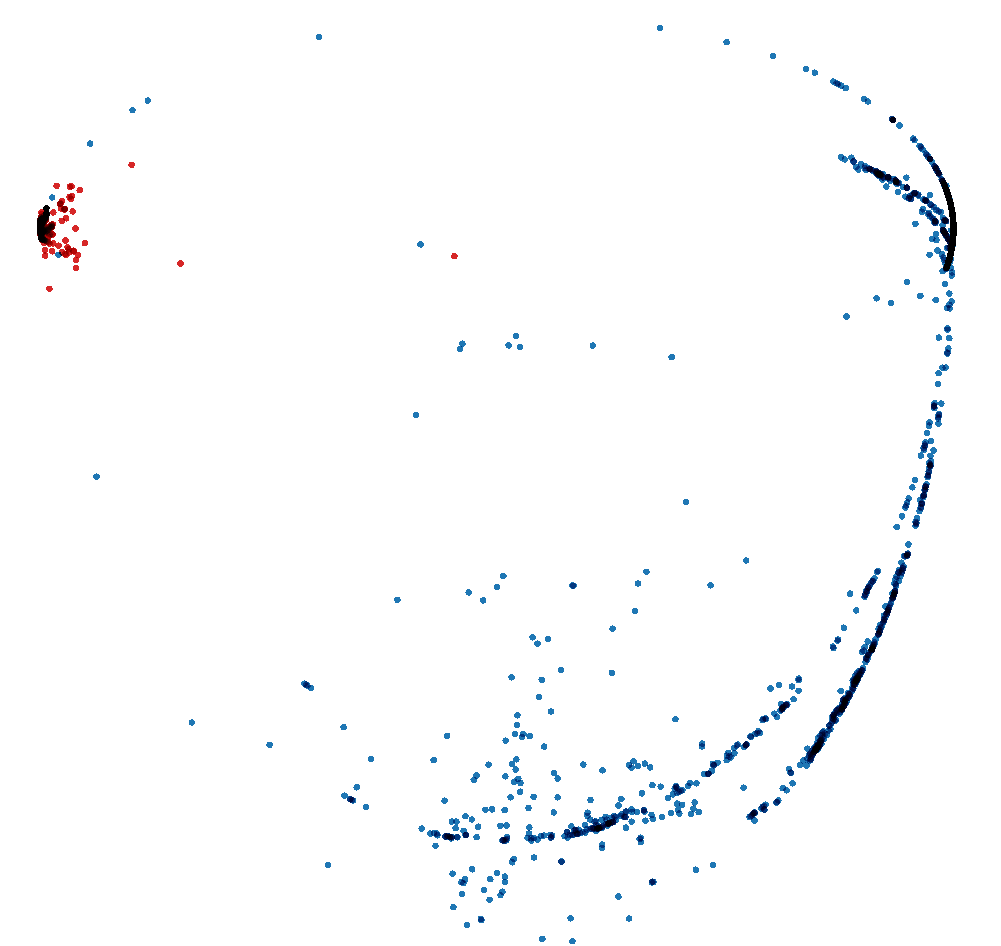}
    \caption{PCA}
  \end{subfigure}
  \hfill 
  \begin{subfigure}[b]{0.32\linewidth}
    \includegraphics[width=\linewidth]{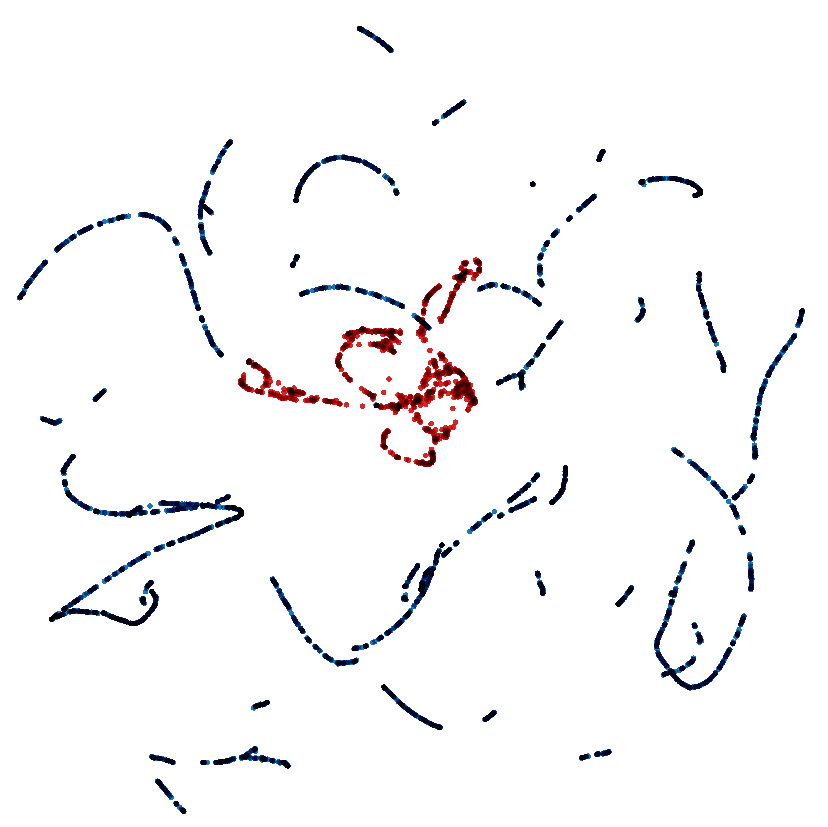}
    \caption{UMAP}
  \end{subfigure}
  \hfill 
  \begin{subfigure}[b]{0.32\linewidth}
    \includegraphics[width=\linewidth]{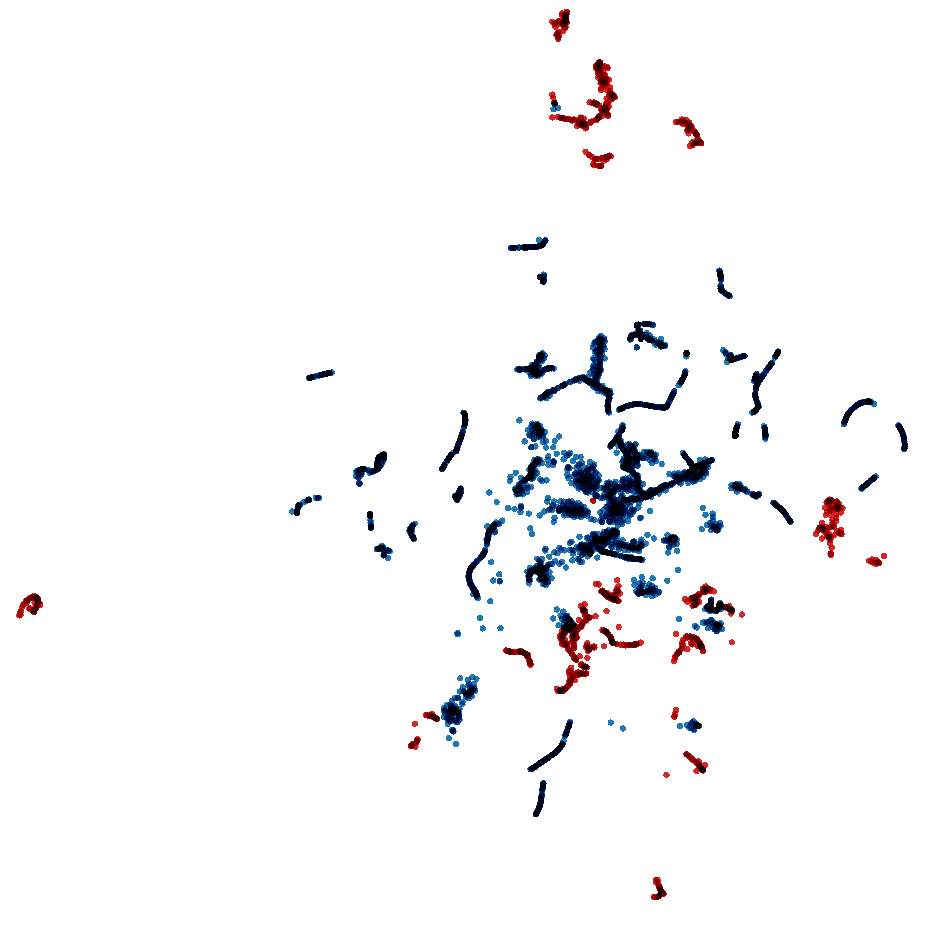}
    \caption{T-SNE}
  \end{subfigure}

    \caption{The visualization of vectors embedding by the embedding neural network using different dimension reduction algorithms.}
    \label{figure_visualization}
\end{figure}

%% file: texts/conclusion.tex
This paper proposed a novel approach to detect anomalies in software execution logs using a Siamese network structure with LSTM layers. We compared the results with the state-of-the-art deep-learning-based methods on the HDFS log for anomaly detection dataset and showed that the proposed method achieves the best results on the aforenamed dataset. We conclude that the ability to achieve state-of-the-art performance is due to the Siamese network as the Feedforward neural network with the same architecture offered a considerably lower F1-score (0.996 vs. 0.973).
Furthermore, we proposed a novel algorithm to generate pairs to train the Siamese network to reduce the training process's computational cost while maintaining accuracy.  We also showed that the Siamese network achieves satisfactory results with smaller and cheaper neural networks as well.
Moreover, we introduced multiple practical advantages of the Siamese network. We assess the robustness of our model to log evolutions. Additionally, we introduced an unsupervised method for log evolution measurement. Finally, we visualize the embedding function's output vectors using dimension reduction algorithms to make the neural network's output more perceptible.

Although we introduced various applications for the Siamese network alongside anomaly detection, there are interesting future investigations. Future works may focus on applying different side applications such as Root Cause Analysis by applying the Siamese network. More computationally cost-efficient neural networks such as CNNs might be applied inside the Siamese neural network to further reduce the computational cost in future studies.

%% file: texts/acknowledgements.tex
This work has been supported by the Academy of Finland (grant IDs 298020 and 328058). Additionally, the authors gratefully acknowledge CSC – IT Center for Science, Finland, for their generous computational resources.